\documentclass[twocolumn]{aastex6}

\newcommand\teff{\mbox{$T_\mathrm{eff}$}}
\usepackage{color}

\begin{document}

\title{The Collapse of the Wien Tail in the Coldest Brown Dwarf? Hubble Space Telescope Near-Infrared Photometry of WISE J085510.83$-$071442.5}

\author{Adam C. Schneider\altaffilmark{1}, Michael C. Cushing\altaffilmark{1}, J. Davy Kirkpatrick\altaffilmark{2}, \& Christopher R. Gelino\altaffilmark{2,3}}  

\altaffiltext{1}{Department of Physics and Astronomy, University of Toledo, 2801 W. Bancroft St., Toledo, OH 43606, USA; Adam.Schneider@Utoledo.edu}
\altaffiltext{2}{Infrared Processing and Analysis Center, MS 100-22, California Institute of Technology, Pasadena, CA 91125, USA}
\altaffiltext{3}{NASA Exoplanet Science Institute, Mail Code 100-22, California Institute of Technology, 770 South Wilson Ave, Pasadena, CA 91125, USA}

\begin{abstract}

We present {\it Hubble Space Telescope (HST)} near-infrared photometry of the coldest known brown dwarf, WISE J085510.83$-$071442.5 (WISE 0855$-$0714).  WISE 0855$-$0714 was observed with the Wide Field Camera 3 (WFC3) aboard {\it HST} using the F105W, F125W, and F160W filters, which approximate the $Y$, $J$, and $H$ near-infrared bands.  WISE 0855$-$0714 is undetected at F105W with a corresponding 2$\sigma$ magnitude limit of $\sim$26.9.  We marginally detect WISE 0855$-$0714 in the F125W images (S/N $\sim$4), with a measured magnitude of 26.41 $\pm$ 0.27, more than a magnitude fainter than the $J-$band magnitude reported by Faherty and coworkers.  WISE J0855$-$0714 is clearly detected in the F160W band, with a magnitude of 23.86 $\pm$ 0.03, the first secure detection of WISE 0855$-$0714 in the near-infrared.  Based on these data, we find that WISE 0855$-$0714 has extremely red F105W$-$F125W and F125W$-$F160W colors relative to other known Y dwarfs.  We find that when compared to the models of Saumon et al.\ and Morley et al., the F105W$-$F125W and F125W$-$F160W colors of WISE 0855$-$0714 cannot be accounted for simultaneously.  These colors likely indicate that we are seeing the collapse of flux on the Wien tail for this extremely cold object.       

\end{abstract}


\section{Introduction}
Without a stable internal energy source, brown dwarfs continuously cool over time.  At their lowest temperatures, their atmospheres are composed primarily of molecules in both the gas phase and solid phase (in the form of condensate clouds) and thus their emergent spectra are sculpted by broad molecular absorption bands of H$_2$O, CH$_4$, and NH$_3$.  The exact chemical composition of their atmospheres evolves as they cool, which gives rise to the smooth variation in spectral morphology that is reflected in the MLTY spectral sequence.

The vast majority of brown dwarfs in the field ($>$1000) were discovered by wide-field surveys such as the Two Micron All Sky Survey and the Sloan Digital Sky Survey (e.g., \citealt{kirk99}, \citealt{haw02}, \citealt{cruz03}, \citealt{chiu06}).  These surveys were adept at identifying the relatively warm L (\teff=2400--1400 K) and T (\teff=1400--700 K) dwarfs because they operated at red-optical (0.7$-$1.0 $\mu$m) and near-infrared (1$-$2.5 $\mu$m) wavelengths.  The evolution of a brown dwarf through the M$\rightarrow$T sequence is controlled by two main processes: 1) the formation of dust and 2) a shift in the carbon chemistry (see however, \cite{trem16} for an alternative explanation).  The M/L transition is marked by the formation of dust clouds that remove refractory elements (TiO, VO, Fe, Si, Al) from the gas phase and generate their own continuum opacity.  The L/T transition is signaled by a sudden loss of this opacity (by some as-yet unknown mechanism) and a shift in the carbon chemistry from being CO-dominated to CH$_4$-dominated. Our confidence that the basic physics for L and T  dwarfs is well understood is buoyed by the fact that atmospheric models do a reasonably good job of matching their observed spectra from the red-optical to the mid-infrared (e.g., \citealt{cush08}, \citealt{ste09}).

Identifying even cooler brown dwarfs required moving to mid-infrared (2.5$-$5 $\mu$m) wavelengths and indeed the {\it Spitzer Space Telescope} and the Wide-field Infrared Survey Explorer {\it (WISE)} have discovered twenty-two of the twenty four brown dwarfs known with effective temperatures less than $\sim$500 K (\citealt{cush11}, \citealt{luh11}, \citealt{kirk12}, \citealt{tin12}, \citealt{kirk13}, \citealt{cush14}, \citealt{pin14}, \citealt{luh14a}, \citealt{dup15}, and \citealt{schneid15}), the  exceptions being CFBDSIR J1458$+$1013 \citep{liu11} and WISE J1217$+$1626B, a companion to a {\it WISE } identified late-T dwarf (\citealt{liu12}, \citealt{leg15}).  These brown dwarfs, which populate the Y spectral class (\citealt{cush11}, \citealt{kirk12}), have proven much harder to understand primarily because of their paucity and intrinsic faintness ($M_J \gtrsim 20$).  In particular, unlike in the case of the L and T dwarfs, model atmospheres struggle to simultaneously account for near- and mid-infrared observations of Y dwarfs \citep{schneid15} which suggests that they are either missing or incorrectly modeling important physics.  Yet it is exactly these cold brown dwarfs that provide the best constraints on the brown dwarf mass function \citep{burg04} and the ultracool model atmospheres required to interpret observations of the gas giant exoplanets discovered with the next-generation adaptive optics instruments (e.g.\ the Gemini Planet Imager and VLT's SPHERE).  

WISE J085510.83$-$071442.5 (hereafter WISE 0855$-$0714) was independently identified using multi-epoch {\it WISE} observations by \cite{luh14} and \cite{kirk14} as a high proper motion object, but it was \cite{luh14a} who measured a distance of only 2 pc, securing it as the fourth closest system to the Sun.  The extremely red colors of [3.6]$-$[4.5] = 3.55 mag\footnote{where [3.6] and [4.5] refer to channel 1 and channel 2 of the InfraRed Array Camera (IRAC; \citealt{faz04}) aboard the {\it Spitzer Space Telescope}} and J$-$[4.5] $>$ 9.11 mag (\citealt{luh14b}, \citealt{fah14}), and faint absolute magnitude of $M_{\rm [4.5]}$ $\sim$17.1 mag \citep{luh14b} make WISE 0855 the reddest, faintest, and coldest (\teff\ $\sim$250 K) brown dwarf known.  To date, there have been four published efforts to image WISE 0855$-$0714 in the near-infrared, resulting in three upper limits ($Y$ $>$ 24.4 mag, \citealt{bea14}; $z_{\rm AB}$ $>$ 24.8 mag, \citealt{kop14}; and $H$ $>$ 22.7 mag, \citealt{wri14}) and one tentative detection ($J_{\rm MKO}$ = 24.8$^{+0.53}_{-0.35}$ mag at 2.6$\sigma$; \citealt{fah14} (see Table 1)).  WISE 0855$-$0714 is therefore the only known Y dwarf without a secure detection in the near-infrared.  Since WISE 0855$-$0714 is simply too faint obtain a near-infrared spectrum, even with the {\it Hubble Space Telescope (HST)}, we obtained {\it HST} near-infrared images of WISE 0855$-$0714 to measure its near-infrared spectral energy distribution.

\section{HST/WFC3 Observations}

We observed WISE 0855$-$0714 with the F105W ($\lambda_p$ = 1055.2 nm), F125W ($\lambda_p$ = 1248.6 nm), and F160W ($\lambda_p$ = 1536.9 nm) filters of Wide Field Camera 3 (WFC3; \citealt{kim08}) aboard {\it HST}, where $\lambda_p$ is the ``pivot wavelength'' (see \citealt{tok05}).  These filters coincide roughly with the $Y$, $J$, and $H$ photometric bands.  The F105W and F125W observations took place on UT 2016 Mar 1 with total exposure times of 3835s and 3635s, respectively.  The F160W observations took place on UT 2016 Mar 15 and UT 2016 Mar 27, with a total exposure time of 25920s.  

\begin{deluxetable}{lccc}
\tablecaption{WISE J085510.83$-$071442.5 Photometry}
\tablehead{
\colhead{Parameter} & \colhead{Value} & \colhead{Ref.}}
\startdata
z & $>$24.8\tablenotemark{b} mag & 2\\
Y & $>$24.4\tablenotemark{c} mag & 3\\
F105W &  $>$26.9\tablenotemark{a} mag & 1\\
J & 25.0$^{+0.53}_{-0.35}$ (or $>$ 24.0\tablenotemark{d}) mag & 4\\
F125W &  26.41 $\pm$ 0.27 mag & 1\\
H & $>$22.7\tablenotemark{c} mag & 5\\
F160W &  23.86 $\pm$ 0.03 mag & 1\\
K & $>$18.6\tablenotemark{c} mag & 6\\
W1 & 17.819 $\pm$ 0.327 mag & 5\\
W2 & 14.02 $\pm$ 0.05 mag & 5\\
$[3.6]$ & 17.44 $\pm$ 0.05 mag & 6 \\
$[4.5]$ & 13.89 $\pm$ 0.02 mag & 6\\
\enddata
\tablenotetext{a}{2$\sigma$ upper limit (Section 2).}
\tablenotetext{b}{AB magnitude; S/N $<$ 3}
\tablenotetext{c}{S/N $<$ 3}
\tablenotetext{d}{S/N $<$ 5}
\tablerefs{ (1) This work; (2) \cite{kop14}; (3) \cite{bea14}; (4) \cite{fah14}; (5) \cite{wri14}; (6) \cite{luh14a} }

\end{deluxetable}

\begin{figure*}
\plotone{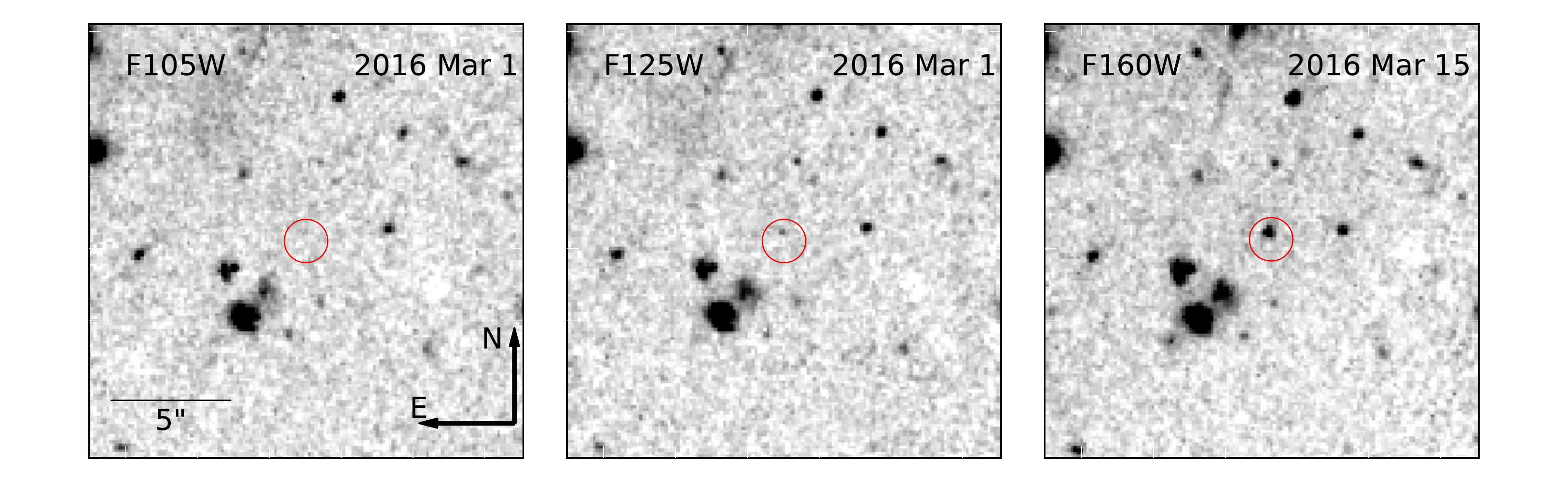}
\caption{F105W, F125W, and F160W images centered on the expected position of WISE 0855$-$0714 on UT 2016 Mar 1.  The red circle indicates the expected position of WISE 0855$-$0714 based on the parallax and proper motion in \cite{luh14b} and the epoch of the observations.  North is up and east is left, and the scale is shown in the bottom left corner of the F105W image.  
}  
\end{figure*}

All images for each filter were aligned and then combined with the {\it tweakreg} and {\it astrodrizzle} routines available in the AstroDrizzle software \citep{gonz12}.  However, there were several individual exposures for which the World Coordinate System (WCS) was in error during the first visit for the F160W observations.  These few frames were omitted when constructing our F160W images.  This has no affect on any resulting science conclusions because WISE 0855$-$0714 is well detected at F160W. We created drizzled images for all three visits individually.  Even though some frames were excluded from the first visit, the total exposure time was the same as that of the other visits, 7770 seconds.  The photometry of WISE 0855$-$0714 from each of the F160W images was consistent to within 0.03 magnitudes (less than 1$\sigma$).  Thirty square arcsecond cut-outs of the combined F105W, F125W, and F160W images around the position of WISE 0855$-$0714 are shown in Figure 1.  Any differences between expected positions and {\it HST} positions are due to either proper motion and parallax uncertainties, positional uncertainties associated with the IRAC coordinates from which the expected positions were calculated, or offsets of the {\it HST} WCS.  We do not detect WISE 0855$-$0714 in the F105W image at its expected position.  While WISE 0855$-$0714 does appear in the F125W image, it is very faint, with a signal-to-noise ratio (S/N) of $\sim$4.  We determine F125W and F160W magnitudes following the method of \cite{schneid15}, whereby we place random apertures around the image in order to determine the background flux.  This process is used because the noise in adjacent pixels becomes correlated as part of the drizzling process, thereby making the common practice of using a sky annulus to determine the noise of the background unsuitable.  The F160W image in Figure 1 and the F160W photometry in Table 1 come from the first visit.   

For the F105W image, we measure an upper limit following the method in the {\it WISE} explanatory supplement\footnote{http://wise2.ipac.caltech.edu/docs/release/prelim/expsup/sec4\_3c.html\#ul2}, where we take the flux measurement plus two times the measurement uncertainty as the 95\% confidence upper limit.  The aperture for this measurement is placed on the F105W image at the position of WISE 0855$-$0714 in the F125W image because both sets of observations were executed on the same day.  {\it HST} photometry from this program, as well as a summary of all previously published photometry for WISE 0855$-$0714 is provided in Table 1.  All photometry is on the Vega system unless otherwise noted.

\section{Analysis}

\begin{figure*}
\plotone{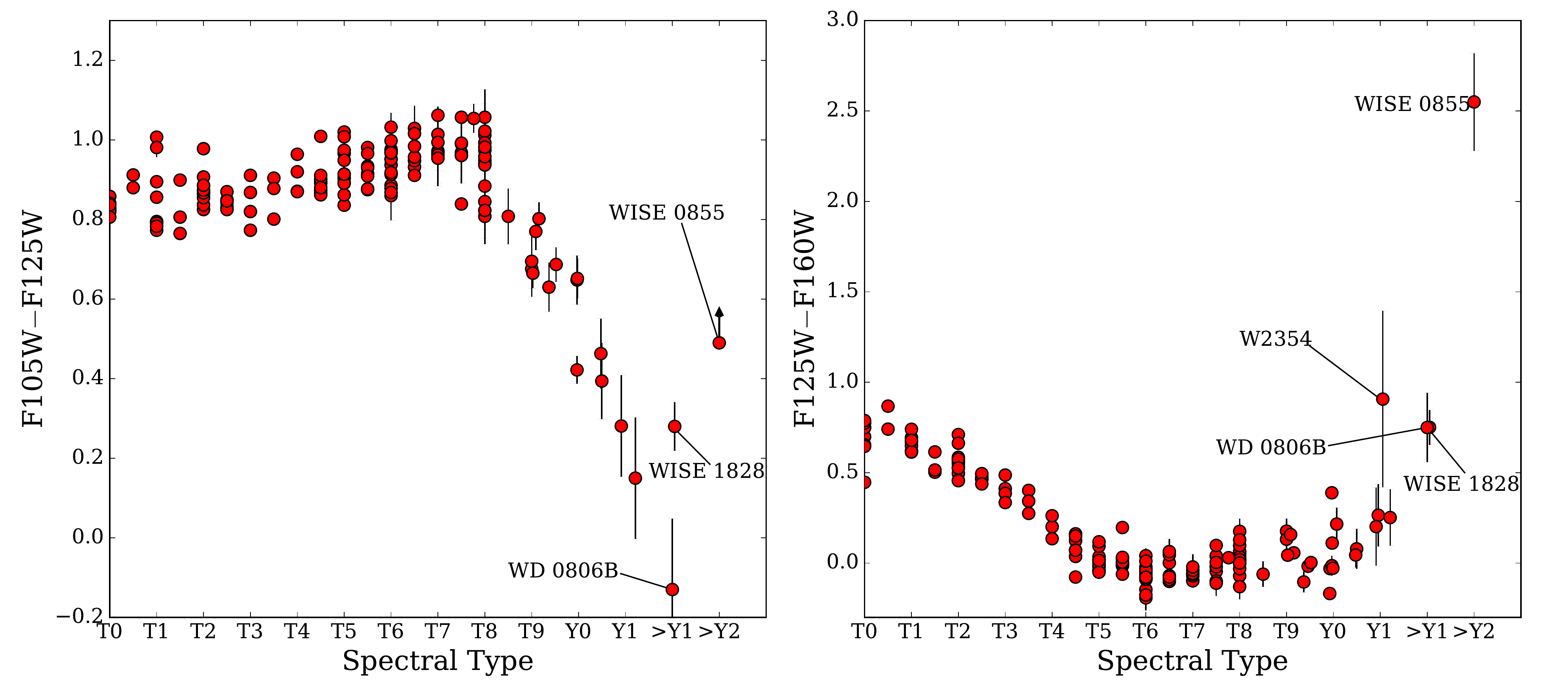}
\caption{{\it HST} F105W$-$F125W and F125W$-$F160W colors as a function of spectral type for T and Y dwarfs.  }  
\end{figure*}

\subsection{Colors Versus Spectral Type}
With near-infrared photometry in hand, we can now place WISE 0855$-$0714 in a broader context by comparing its near-infrared colors to those of known T and Y dwarfs.  Figure 2 shows the F105W$-$F125W and F125W$-$F160W colors of a sample of T and Y dwarfs and WISE 0855$-$0714.  All T dwarf colors are found synthetically from {\it HST} system throughput tables\footnote{http://stsci.edu/hst/wfc3/ins\_performance/throughputs} and near-infrared spectra found in the SpeX Prism Library\footnote{http://pono.ucsd.edu/$\sim$adam/browndwarfs/spexprism/} \citep{burg14}.  Y dwarf colors are determined synthetically from the {\it HST} spectroscopic sample of \cite{schneid15}.  Because there are very few late-T dwarfs ($>$T7) in the SpeX Prism Library and in the {\it HST} sample of \cite{schneid15}, we supplement these two datasets with late-T spectra from \cite{kirk11}.  Also shown is the near-infrared photometry of two other cold brown dwarfs with estimated effective temperatures less than 300 K; WD 0806$-$661B \citep{luh11} and WISE 182831.08$+$265037.6 \citep{cush11}\footnote{Note, however, that an independent estimate of the effective temperature of WISE 182831.08$+$265037.6 based on its bolometric luminosity of 520$^{+60}_{-50}$ K \citep{dup13} implies that WISE 182831.08$+$265037.6 is actually much warmer.  It has been suggested that WISE 182831.08$+$265037.6 is in fact an unresolved binary, which would help explain its unusual luminosity \citep{leg13}.}. The photometry for WISE 182831.08$+$265037.6 is determined synthetically from its spectrum (Cushing et al., in preparation), while the photometry for WD0806$-$661B is determined directly from {\it HST} images (Gelino et al., in preparation).  Because a spectral type of WISE 0855$-$0714 is currently unknown, we label it as $>$Y2.  Similarly, WISE 182831.08$+$265037.6 and WD0806$-$661B are denoted as $>$Y1.

The severe turn towards the blue of the F105W$-$F125W color at the T/Y boundary is typically ascribed to the loss of the K I and Na I opacity in the red optical as these atoms condense into KCl and Na$_2$S (\citealt{lod99}, \citealt{burn10}).   A similar trend is seen in the $z-J$ \citep{lod13} and $Y-J$ (\citealt{burn10}, \citealt{liu12}, \citealt{leg13}, \citealt{leg15}, \citealt{schneid15}) colors of Y dwarfs.  Note, however, that the measured $Y-J$ colors for Y dwarfs are typically much bluer than models that incorporate this loss predict (e.g., \citealt{mor14}), suggesting that there may be other factors affecting the flux around 1 micron (see \citealt{liu12}).

At even lower temperatures, models predict this trend turns back towards the red with the precipitous loss of the Wien tail \citep{bur03}. The position of WD 0806$-$661B implies the blueward trend continues. However, WISE 182831.08$+$265037.6 may be showing the first hints of this turnaround. Even though WISE 0855$-$0714 was undetected in the F105W image, the F105W$-$F125W color limit is redder than those of every other brown dwarf with spectral type $\geq$ Y1, showing that the blueward trend must turn back toward the red somewhere between the Y1 dwarfs, WD0806$-$661B, WISE 182831.08$+$265037.6, and WISE 0855$-$0714. 

The F125W$-$F160W colors plateau for T7 to Y0 type objects, but there have been hints that this trend also turns back to the red (e.g., \citealt{cush11}, \citealt{kirk12}, \citealt{leg13}, \citealt{cush14}, \citealt{leg15}, \citealt{schneid15}).  Interestingly, atmospheric models predict a turn to the blue around 300 K, since the water clouds that form at these temperatures absorb light in F160W but not in F125W \citep{mor14}.  We do not see this blueward trend.  Instead, the inclusion of WISE 182831.08$+$265037.6, WD 0806$-$661B, and WISE 0855$-$0714 on this diagram shows definitively that the F125W$-$F160W color turns toward the red for the coldest brown dwarfs.  In summary, both the F105W$-$F125W color limit and F125W$-$F160W are extremely red compared to known Y dwarfs, and can likely be attributed to the collapse of the Wien tail at the low temperature of WISE 0855$-$0714. 

\subsection{Color-Magnitude Diagrams}

In order to properly compare the {\it HST} photometry of WISE 0855$-$0714 with that of other brown dwarfs and low-temperature models on color-magnitude diagrams, we need to determine {\it HST} photometry for those samples.  For the models, colors are found synthetically from the model spectra.  To obtain absolute F125W magnitudes, we measure the $J_{\rm MKO}-$F125W color for each model spectrum synthetically, then apply these offsets to the absolute $J_{\rm MKO}$ provided for each model\footnote{http://www.ucolick.org/$\sim$cmorley/cmorley/Models.html}.  Colors and absolute magnitudes for known brown dwarfs and low-mass stars are found in a similar fashion: we first determine colors synthetically from spectra available in the SpeX Prism Library, \cite{kirk11}, and \cite{schneid15}.  For the {\it HST} spectra, we measure F125W and F160W magnitudes directly since the spectra are absolutely flux calibrated.  For spectra from the SpeX Prism Library and \cite{kirk11}, we first compute the $J_{\rm MKO}-$F125W color synthetically from the spectrum, and determine F125W magnitudes from measured $J_{\rm MKO}$ values.  The majority of $J_{\rm MKO}$ and parallax measurements come from the Database of Ultracool Parallaxes maintained by Trent Dupuy\footnote{http://www.as.utexas.edu/$\sim$tdupuy/plx/Database\_of\_Ultracool\_Parallaxes.html} \citep{dup12}.  Additional photometry and parallaxes were taken from \cite{kirk11}, \cite{scholz12}, \cite{mace13}, \cite{marsh13}, \cite{beich14}, and \cite{tin14}.  The average values of $J_{\rm MKO}-$F125W and $H_{\rm MKO}-$F160W for known Y dwarfs are $-$0.69 and $-$0.16 mag, respectively.

\begin{figure*}
\plotone{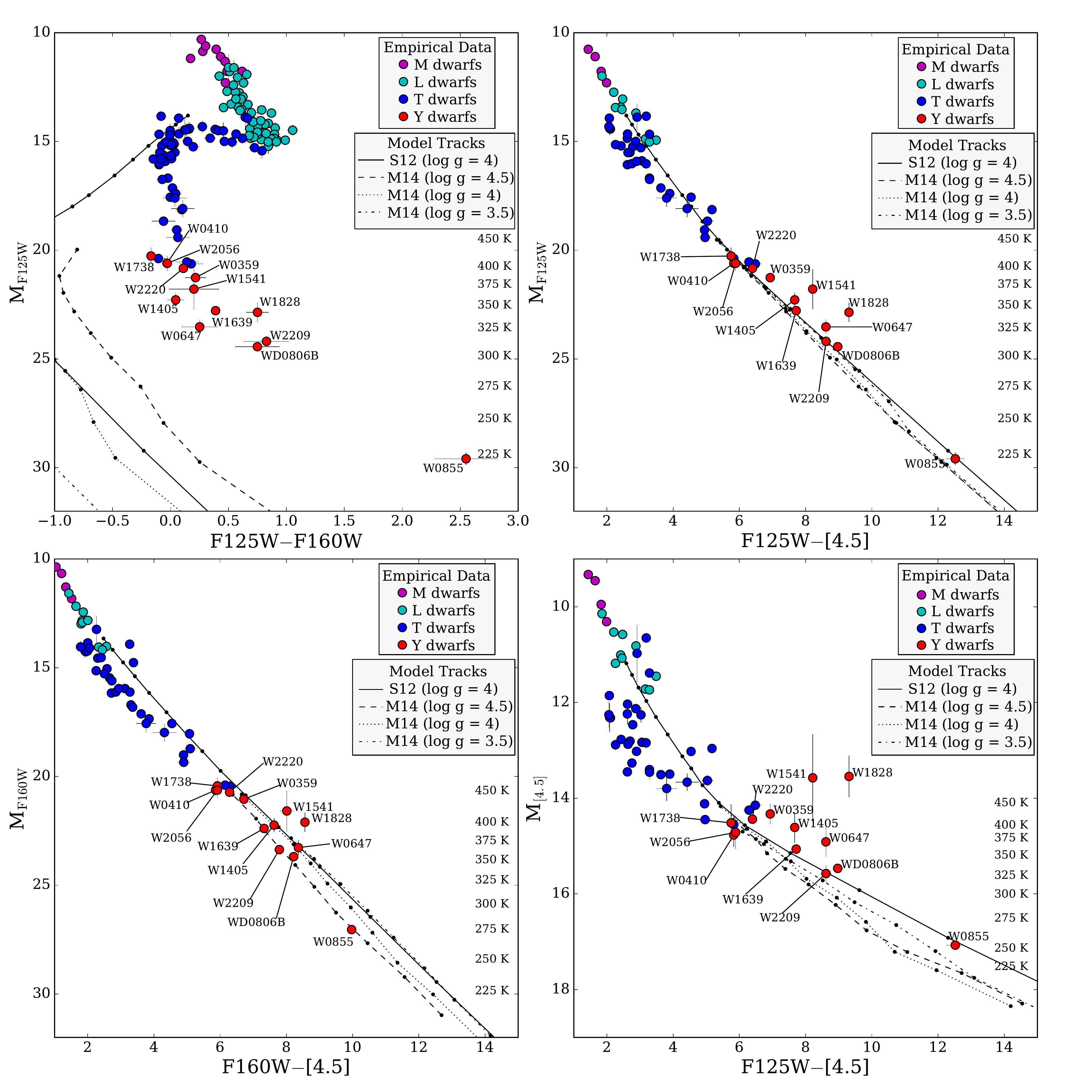}
\caption{Color magnitude diagrams for WISE 0855$-$0714 and M (purple), L (cyan), T (blue), and Y (red) dwarfs.  All Y dwarfs are individually labeled.  S12 and M14 refer to \cite{sau12} and \cite{mor14}, respectively.  The temperatures listed along the right-hand side of each panel correspond to the \cite{mor14} log g = 4, $f_{\rm sed}$ = 5 model tracks (dotted line).}  
\end{figure*}

Previous studies have shown that $J-H$ colors of the coldest brown dwarfs are not well reproduced by low temperature models (e.g., \citealt{leg15}, \citealt{schneid15}).  Models generally predict much bluer colors for a given absolute magnitude than what is measured.  The position of WISE 0855$-$0714 in the F125W$-$F160W color-magnitude diagram in the top left panel of Figure 3 shows this issue extends down to the temperature of WISE 0855$-$0714.  We use the WISE 0855$-$0714 parallax from \cite{luh14b} for this figure.  We include in Figure 3 cloud-free models from \cite{sau12} and models that include sulfide and water clouds and are 50\% cloudy \citep{mor14}.  The lowest temperature 100\% cloudy models of \cite{mor12} are 400 K, much warmer than \teff\ estimates for WISE 0855$-$0714, so are not considered here.  All models used have solar metallicities and are in chemical equilibrium.  For the 50\% cloudy models of \cite{mor14}, very little variation is seen in the photometry for different values of sedimentation efficiency ($f_{\rm sed}$), so we choose to only compare to $f_{\rm sed}$ = 5 models.  We also chose only to employ models with surface gravities consistent with the evolutionary models of \cite{sau08}.

As seen in the figure, there is no single model down to \teff\ $\sim$225 K that reaches a F125W$-$F160W color as red as the color of WISE 0855$-$0714.  Note that the log g = 3.5 and much of the log g = 4 model tracks have bluer colors than the limits of the figure.  The F125W$-$F160W color of the \teff\ = 225 K, log g = 4.5, $f_{\rm sed}$ = 5 model of \cite{mor14} comes the closest to WISE 0855$-$0714.  However, that model has a F105W$-$F125W color of $\sim$0.1 mag, which is clearly ruled out by our F105W upper limit (see Figure 2).  

\cite{leg16} showed that the cloud-free models of \cite{trem15} are more consistent with the $J-H$ colors of Y dwarfs than those of \cite{sau12} (see their Figure 4).  Applying our $J_{\rm MKO}-$F125W and $H_{\rm MKO}-$F160W offsets to WISE 0855$-$0714 we find $M_{\rm J}$ $\approx$ 28.9 mag and $J-H$ $\approx$ 2.0 mag.  The only model track in Figure 4 of \cite{leg16} that extends down to an $M_{\rm J}$ value of 29 mag has a log g = 4.5 and is computed with equilibrium chemistry.  While that model track extends beyond the boundaries of the figure, we estimate its $J-H$ color to be $\sim$1.5, much closer to WISE 0855$-$0714's color of $J-H$ $\approx$ 2.0 than the color predicted by the \cite{sau12} models of $J-H$ $\approx$ 0. 

The top and bottom right panels of Figure 3 show the absolute F125W and [4.5] magnitudes as a function of F125W$-$[4.5] color.  Model F125W$-$[4.5] colors reproduce the measured colors of cold brown dwarfs much better than the F125W$-$F160W model colors.  When compared to the log g = 4 \cite{mor14} models that extend down to 200 K, the position of WISE 0855$-$0714 on these figure suggests a temperatures between $\sim$225 K and $\sim$250 K, similar to previous estimates (\citealt{bea14}, \citealt{fah14}, \citealt{kop14}, \citealt{luh14a}, \citealt{luh14b}).

We also show the F160W absolute magnitude as a function of F160W$-$[4.5] color in the bottom left panel of Figure 3.  This diagram is notable because the F160W and [4.5] bands have the smallest photometric uncertainties of all the measurements in Table 1.  In this diagram, WISE 0855$-$0714 lies very close to the log g = 4.5 model track, with a temperature between 250 and 275 K, slightly warmer than estimates using other photometric bands.  It should not be too surprising that the temperature estimate from the F160W magnitude is warmer than that from the F125W magnitude, considering that the F160W magnitude of WISE 0855$-$0714 is so much brighter.    

\section{Discussion}

\cite{fah14}, using their 2.6$\sigma$ $J-$band detection of WISE 0855$-$0714, conclude that WISE 0855$-$0714 likely contains water clouds within its atmosphere based on its position in a diagram comparing its absolute {\it WISE} channel 2 (W2) magnitudes and $J-$W2 color to those from models.  However, using an updated parallax measurement and the $J-$band measurement of \cite{fah14}, \cite{luh14b} find that the absolute $[4.5]$ magnitude and the $J-$[4.5] color were consistent with \cite{sau12} cloudless models with an atmosphere whose carbon and nitrogen chemistry is out of equilibrium due to vertical mixing in the atmosphere. Our new position for WISE 0855$-$0714 in the M$_{\rm [4.5]}$ vs. F125W$-$[4.5] color-magnitude diagram (bottom right panel of Figure 3) does little to clear up whether it better matches cloud-free, water-cloud, or non-equilibrium chemistry models because we do not have non-equilibrium models.  Nevertheless, the fact that its position agrees extremely well with the \cite{mor14} water-cloud models in some diagrams in Figure 3, and not as well in others, suggests that drawing conclusions from a single color-magnitude diagram when considering competing models may be premature.  Instead, a more comprehensive view of its full spectral energy distribution is likely needed.

Recent work has shown that non-equilibrium chemistry driven by vertical mixing is likely important in cold brown dwarf atmospheres, and may play a large role in shaping their final spectroscopic shapes (e.g., \citealt{schneid15}, \citealt{trem15}, \citealt{leg16}).  Non-equilibrium chemistry in the atmosphere of WISE 0855$-$0714 likely makes a large contribution to the discrepancies between models and observations (e.g., the F125W$-$F160 color differences).  A larger suite of models including the effects of non-equilibrium chemistry and cloud formation that extends down to \teff\ values of $\sim$200 K will be necessary to draw any further conclusions of the atmospheric properties of this extremely cold object.

\section{Acknowledgments}

We thank the anonymous referee whose comments improved the clarity of this paper.  We thank Mark Marley, Didier Saumon, and Caroline Morley for fruitful discussions and for graciously making their models publicly available online. This work is Based on observations made with the NASA/ESA Hubble Space Telescope, obtained at the Space Telescope Science Institute, which is operated by the Association of Universities for Research in Astronomy, Inc., under NASA contract NAS 5-26555. These observations are associated with program \#14233.  Support for program \#14233 was provided by NASA through a grant from the Space Telescope Science Institute, which is operated by the Association of Universities for Research in Astronomy, Inc., under NASA contract NAS 5-26555.  This publication makes use of data products from the {\it Wide-field Infrared Survey Explorer}, which is a joint project of the University of California, Los Angeles, and the Jet Propulsion Laboratory/California Institute of Technology.  This research has benefitted from the SpeX Prism Spectral Libraries, maintained by Adam Burgasser at http://pono.ucsd.edu/$\sim$adam/browndwarfs/spexprism.


\begin{thebibliography}{}
\bibitem[Beam{\'{\i}}n et al.(2014)]{bea14} Beam{\'{\i}}n, J.~C., Ivanov, V.~D., Bayo, A., et al.\ 2014, \aap, 570, L8 
\bibitem[Beichman et al.(2014)]{beich14} Beichman, C., Gelino, C.~R., Kirkpatrick, J.~D., et al.\ 2014, \apj, 783, 68 
\bibitem[Burgasser(2004)]{burg04} Burgasser, A.~J.\ 2004, \apjs, 155, 191 
\bibitem[Burgasser(2014)]{burg14} Burgasser, A.~J.\ 2014, ASInC, 11, 7 
\bibitem[Burningham et al.(2010)]{burn10} Burningham, B., Pinfield, D.~J., Lucas, P.~W., et al.\ 2010, \mnras, 406, 1885 
\bibitem[Burrows et al.(2003)]{bur03} Burrows, A., Sudarsky, D., \& Lunine, J.~I.\ 2003, \apj, 596, 587 
\bibitem[Chiu et al.(2006)]{chiu06} Chiu, K., Fan, X., Leggett, S.~K., et al.\ 2006, \aj, 131, 2722 
\bibitem[Cruz et al.(2003)]{cruz03} Cruz, K.~L., Reid, I.~N., Liebert, J., Kirkpatrick, J.~D., \& Lowrance, P.~J.\ 2003, \aj, 126, 2421 
\bibitem[Cushing et al.(2008)]{cush08} Cushing, M.~C., Marley, M.~S., Saumon, D., et al.\ 2008, \apj, 678, 1372-1395 
\bibitem[Cushing et al.(2011)]{cush11} Cushing, M.~C., Kirkpatrick, J.~D., Gelino, C.~R., et al.\ 2011, \apj, 743, 50
\bibitem[Cushing et al.(2014)]{cush14} Cushing, M.~C., Kirkpatrick, J.~D., Gelino, C.~R., et al.\ 2014, \aj, 147, 113
\bibitem[Dupuy \& Liu(2012)]{dup12} Dupuy, T.~J., \& Liu, M.~C.\ 2012, \apjs, 201, 19 
\bibitem[Dupuy \& Kraus(2013)]{dup13} Dupuy, T.~J., \& Kraus, A.~L.\ 2013, Science, 341, 1492 
\bibitem[Dupuy et al.(2015)]{dup15} Dupuy, T.~J., Liu, M.~C., \& Leggett, S.~K.\ 2015, \apj, 803, 102 
\bibitem[Faherty et al.(2014)]{fah14} Faherty, J.~K., Tinney, C.~G., Skemer, A., \& Monson, A.~J.\ 2014, \apjl, 793, L16 
\bibitem[Fazio et al.(2004)]{faz04} Fazio, G.~G., Hora, J.~L., Allen, L.~E., et al.\ 2004, \apjs, 154, 10 
\bibitem[Gonzaga et al.(2012)]{gonz12} Gonzaga, S., \& et al.\ 2012, The DrizzlePac Handbook, HST Data Handbook, (Baltimore, STScI)
\bibitem[Hawley et al.(2002)]{haw02} Hawley, S.~L., Covey, K.~R., Knapp, G.~R., et al.\ 2002, \aj, 123, 3409 
\bibitem[Kimble et al.(2008)]{kim08} Kimble, R.~A., MacKenty, J.~W., O'Connell, R.~W., \& Townsend, J.~A.\ 2008, \procspie, 7010, 43
\bibitem[Kirkpatrick et al.(1999)]{kirk99} Kirkpatrick, J.~D., Reid, I.~N., Liebert, J., et al.\ 1999, \apj, 519, 802 
\bibitem[Kirkpatrick et al.(2011)]{kirk11} Kirkpatrick, J.~D., Cushing, M.~C., Gelino, C.~R., et al.\ 2011, \apjs, 197, 19 
\bibitem[Kirkpatrick et al.(2012)]{kirk12} Kirkpatrick, J.~D., Gelino, C.~R., Cushing, M.~C., et al.\ 2012, \apj, 753, 156
\bibitem[Kirkpatrick et al.(2013)]{kirk13} Kirkpatrick, J.~D., Cushing, M.~C., Gelino, C.~R., et al.\ 2013, \apj, 776, 128
\bibitem[Kirkpatrick et al.(2014)]{kirk14} Kirkpatrick, J.~D., Schneider, A., Fajardo-Acosta, S., et al.\ 2014, \apj, 783, 122 
\bibitem[Kopytova et al.(2014)]{kop14} Kopytova, T.~G., Crossfield, I.~J.~M., Deacon, N.~R., et al.\ 2014, \apj, 797, 3 
\bibitem[Leggett et al.(2013)]{leg13} Leggett, S.~K., Morley, C.~V., Marley, M.~S., et al.\ 2013, \apj, 763, 130 
\bibitem[Leggett et al.(2015)]{leg15} Leggett, S.~K., Morley, C.~V., Marley, M.~S., \& Saumon, D.\ 2015, \apj, 799, 37 
\bibitem[Leggett et al.(2016)]{leg16} Leggett, S.~K., Tremblin, P., Saumon, D., et al.\ 2016, arXiv:1603.09400 
\bibitem[Liu et al.(2011)]{liu11} Liu, M.~C., Delorme, P., Dupuy, T.~J., et al.\ 2011, \apj, 740, 108 
\bibitem[Liu et al.(2012)]{liu12} Liu, M.~C., Dupuy, T.~J., Bowler, B.~P., Leggett, S.~K., \& Best, W.~M.~J.\ 2012, \apj, 758, 57 
\bibitem[Lodders(1999)]{lod99} Lodders, K.\ 1999, \apj, 519, 793 
\bibitem[Lodieu et al.(2013)]{lod13} Lodieu, N., B{\'e}jar, V.~J.~S., \& Rebolo, R.\ 2013, \aap, 550, L2 
\bibitem[Luhman et al.(2011)]{luh11} Luhman, K.~L., Burgasser, A.~J., \& Bochanski, J.~J.\ 2011, \apjl, 730, L9 
\bibitem[Luhman(2014a)]{luh14} Luhman, K.~L.\ 2014a, \apj, 781, 4 
\bibitem[Luhman(2014b)]{luh14a} Luhman, K.~L.\ 2014b, \apjl, 786, L18 
\bibitem[Luhman \& Esplin(2014)]{luh14b} Luhman, K.~L., \& Esplin, T.~L.\ 2014, \apj, 796, 6
\bibitem[Mace et al.(2013)]{mace13} Mace, G.~N., Kirkpatrick, J.~D., Cushing, M.~C., et al.\ 2013, \apjs, 205, 6 
\bibitem[Marsh et al.(2013)]{marsh13} Marsh, K.~A., Wright, E.~L., Kirkpatrick, J.~D., et al.\ 2013, \apj, 762, 119 
\bibitem[Morley et al.(2012)]{mor12} Morley, C.~V., Fortney, J.~J., Marley, M.~S., et al.\ 2012, \apj, 756, 172 
\bibitem[Morley et al.(2014)]{mor14} Morley, C.~V., Marley, M.~S., Fortney, J.~J., et al.\ 2014, \apj, 787, 78 
\bibitem[Pinfield et al.(2014)]{pin14} Pinfield, D.~J., Gromadzki, M., Leggett, S.~K., et al.\ 2014, \mnras, 444, 1931 
\bibitem[Saumon \& Marley(2008)]{sau08} Saumon, D., \& Marley, M.~S.\ 2008, \apj, 689, 1327 
\bibitem[Saumon et al.(2012)]{sau12} Saumon, D., Marley, M.~S., Abel, M., Frommhold, L., \& Freedman, R.~S.\ 2012, \apj, 750, 74
\bibitem[Schneider et al.(2015)]{schneid15} Schneider, A.~C., Cushing, M.~C., Kirkpatrick, J.~D., et al.\ 2015, \apj, 804, 92 
\bibitem[Scholz et al.(2012)]{scholz12} Scholz, R.-D., Bihain, G., Schnurr, O., \& Storm, J.\ 2012, \aap, 541, A163 
\bibitem[Stephens et al.(2009)]{ste09} Stephens, D.~C., Leggett, S.~K., Cushing, M.~C., et al.\ 2009, \apj, 702, 154 
\bibitem[Tinney et al.(2012)]{tin12} Tinney, C.~G., Faherty, J.~K., Kirkpatrick, J.~D., et al.\ 2012, \apj, 759, 60
\bibitem[Tinney et al.(2014)]{tin14} Tinney, C.~G., Faherty, J.~K., Kirkpatrick, J.~D., et al.\ 2014, \apj, 796, 39 
\bibitem[Tokunaga \& Vacca(2005)]{tok05} Tokunaga, A.~T., \& Vacca, W.~D.\ 2005, \pasp, 117, 421 
\bibitem[Tremblin et al.(2015)]{trem15} Tremblin, P., Amundsen, D.~S., Mourier, P., et al.\ 2015, \apjl, 804, L17 
\bibitem[Tremblin et al.(2016)]{trem16} Tremblin, P., Amundsen, D.~S., Chabrier, G., et al.\ 2016, \apjl, 817, L19 
\bibitem[Wright et al.(2014)]{wri14} Wright, E.~L., Mainzer, A., Kirkpatrick, J.~D., et al.\ 2014, \aj, 148, 82 

\end{thebibliography}
\end{document}